\documentclass[final,5p,times,twocolumn]{elsarticle}
\usepackage{amssymb}
\usepackage{lineno}
\usepackage[utf8]{inputenc}
\usepackage[T1]{fontenc}
\usepackage{epsfig}
\usepackage{amsmath}
\usepackage{amsfonts}
\usepackage{amssymb}
\usepackage{color}
\usepackage{graphicx}
\usepackage{url,hyperref}
\usepackage{latexsym}
\usepackage{longtable}
\usepackage{float}
\usepackage{epstopdf}
\usepackage{url}
\journal{Physics Letters B}
\begin{document}
\begin{frontmatter}
\title{On $\gamma$-rigid regime of the Bohr-Mottelson Hamiltonian in the presence of a minimal length}
\author{M. Chabab}
\ead{mchabab@uca.ma}
\author{A. El Batoul}
%\author[label1]{A. El Batoul\corref{cor1}}
\ead{elbatoul.abdelwahed@edu.uca.ma}
\author{A. Lahbas}
\ead{alaaeddine.lahbas@edu.uca.ma}
\author{M. Oulne\corref{cor1}}
\ead{oulne@uca.ma}
\address{High Energy Physics and Astrophysics Laboratory, Department of Physics, Faculty of Sciences Semlalia, Cadi Ayyad University, P.O.B 2390, Marrakesh 40000, Morocco.}
\cortext[cor1]{corresponding author}
\begin{abstract}
A prolate $\gamma$-rigid regime of the Bohr-Mottelson Hamiltonian within the minimal length formalism,  involving an infinite square well like potential  in $\beta$ collective shape variable, is developed and  used to describe the spectra of a variety of vibrational-like nuclei. The  effect of the minimal length on the energy spectrum and the wave function is duly investigated.  Numerical calculations are performed for some nuclei revealing a qualitative agreement with the available experimental data.
\end{abstract}

\end{frontmatter}
\section{Introduction}
 During the last decade, the models based on the concepts of critical point symmetries (CPS) related to shape phase transitions provide a very interesting theoretical framework for studies of nuclear structure phenomena. Actually, this interest has increased even more with the insertion of an additional critical point symmetries. Shape phase transitions  have been first considered  in  the framework of the interacting boson model \cite{Iachello1}, which describes collective states of nuclei in terms of collective bosons of angular momentum zero (s-boson) and two (d-boson) in the context of a U(6) overall symmetry, having a dynamical   U(5) (vibrational), SU(3) (prolate deformed rotational or axial rotor) and O(6) ($\gamma$-unstable) as limiting symmetries.  Another important symmetries called E(5) \cite{Iachello2} and X(5) \cite{Iachello3}, which approximate special solutions of the Bohr-Mottelson model \cite{Bohr} with an infinite-well potential and which were offered for the critical points of the shape phase transitions U(5)$\leftrightarrow$O(6) and U(5)$\leftrightarrow$SU(3) respectively, have been realized by Iachello. Later, a $\gamma$ -rigid (with $\gamma=0$ ) version  of the critical symmetry X(5), called  X(3) have been introduced in \cite{Bonatsos1}. Other models considering the extension of X(3) , such as X(3)-$\beta^{2n}$ (n=1,2) \cite{Budaca1} and X(3)-$\beta^6$ \cite{Budaca2} have also been developed not long ago. Besides, several additional attempts have been done to obtain solutions of the Bohr Hamiltonian with a constant mass parameter \cite{Chabab1,Chabab2} as well as within the deformation dependent mass formalism \cite{Chabab3,Bonatsos2}.\\
Recently, a lot of attention has been attracted by the quantum mechanical problems implying a generalized modified commutation relations which includes a minimal length or Generalized	Uncertainty Principle (GUP). Such an important idea was motivated by noncommutative geometry \cite{Witten,Nathan} in the quantum gravity \cite{QGravity1,QGravity2,QGravity3} and the string theory context \cite{StringTh1,StringTh2,StringTh3}. However, the concept of minimal length can be incorporated in the study of  physical systems by considering the deformed canonical commutation relation,
\begin{equation}
\left[X,P\right]=i\hbar\left(1+\alpha^2P^2\right)
\label{E1}
\end{equation}
here $\alpha$  represents the minimal length parameter (a very small positive parameter). This commutation relation leads to the uncertainty relation
\begin{equation}
\Delta X \Delta P\geq \frac{\hbar}{2}\left(1+\alpha\left(\Delta P\right)^2\right)
\label{E2}
\end{equation}
which implies the existence of a minimal length given by $\left(\Delta X\right)_{min}=\hbar \sqrt{\alpha}$.
It should be noted that, since the elaboration of the fundamental principles of the quantum mechanics with GUP in \cite{GUP1,GUP2,GUP3,GUP4}, a much development, in this direction, has been accomplished in order to study   the effect of the minimal length on  quantum systems as well as on  classical ones. Nevertheless, only  few problems are shown to be solved exactly or  approximately. Among them  one can cite the Schrodinger equation for : the harmonic oscillator \cite{Problem1}, the hydrogen atom \cite{Problem2,Problem3,Problem4,Problem5,Problem6}, the inverse square potential \cite{Problem7}, the scattering problem by Yukawa and Coulomb potentials \cite{Problem8} and  square well potential \cite{Problem9}.
In this Letter,  we  study a $\gamma$-rigid version of the Bohr-Mottelson Hamiltonian, within an infinite square well potential  in $\beta$ collective shape variable   as X(3) model, in the presence of a minimal length. Particularly, we investigate the effect of a minimal length on the physical observables such as energy spectrum and eigenfunctions as well as B(E2) electromagnetic transition rates.
	\label{intro}
	\section{Minimal length formalism}	
The theoretical background of minimal length formalism (MLF)  motivated by a Heisenberg algebra and implying a generalized uncertainty principle (GUP)	has been considered recently in \cite{Problem1,Problem9}. In the framework of this formalism, the generalization of the deformed canonical commutation relation (\ref{E1}) is given by \cite{Problem1,Problem9}
	\begin{equation}
	\left[\hat{X_i},\hat{P_j}\right]=i\hbar\left(\delta_{ij}+\alpha\hat{P}^2\delta_{ij}+\alpha'\hat{P_i}\hat{P_j}\right)
	\label{E3}
	\end{equation}
	where $\alpha'$ is an additional parameter which is of the order of $\alpha$. In this case, the components of the momentum operator commute to one another
	\begin{equation}
	\left[\hat{P_i},\hat{P_j}\right]=0
	\label{E4}
	\end{equation}
	However, the commutator between two position operators is in general different from zero
	\begin{equation}
	\left[\hat{X_i},\hat{X_j}\right]=i\hbar\frac{(2\alpha-\alpha')+(2\alpha+\alpha')\alpha\hat{P}^2}{1+\alpha\hat{P}^2}\left(\hat{P_i}\hat{X_j}-\hat{P_j}\hat{X_i}\right)
	\label{E5}
	\end{equation}
	It is clear that the generalized canonical commutation relation (\ref{E3}) leads to the minimal observable length $(\Delta X_i)_{min}=\hbar\sqrt{3\alpha+\alpha'}$.
	In the same context, we have different representations for the canonical operators $X_i$ and $P_i$. Among these representations, one can cite the momentum space representation \cite{Problem1}:
	\begin{equation}
	\hat{X_i}=i\hbar\left[(1+\alpha p^2)\frac{\partial}{\partial p_i}+\alpha'p_ip_j\frac{\partial}{\partial p_j}\right]+\eta p_i,\ \hat{P_i}=p_i
	\label{E6}
	\end{equation}
	and the position representation given by \cite{Problem3,Problem4}:
	\begin{equation}
	\hat{X_i}=\hat{x_i}+\frac{(2\alpha-\alpha')\left(\hat{p}^2\hat{x_i}+\hat{x_i}\hat{p}^2\right)}{4},\  \hat{P_i}=\hat{p_i}\left(1+\frac{\alpha'}{2}\hat{p}^2\right)
	\label{E7}
	\end{equation}
	where $\hat{x_i}$  and $\hat{p_i}$  are the usual position and momentum operators respectively, which  obey the following
	relations $ \left[\hat{x_i},\hat{p_j}\right]=i\hbar \delta_{ij}$ and $\hat{p}^2=\sum_i \hat{p_i}$. Note that  in the case of $\alpha'=2\alpha$, and for the first order on $\alpha$, the following canonical commutator $ \left[\hat{X_i},\hat{X_j}\right]$ vanishes.  As a consequence, Eq. (\ref{E7}) redruces to
	\begin{equation}
	\hat{X_i}=\hat{x_i},\  \hat{P_i}=\left(1+\alpha\hat{p}^2\right)\hat{p_i}
	\label{E8}
	\end{equation}
	In addition, we can interpret $p_i$ and $P_i$ shown in Eq. (\ref{E8}) according to string theory:
	 $p_i$ is the momentum operator at low energies  and $P_i$ is the momentum operator at high energies. Moreover, $p$ is the magnitude of the $p_i$ vector.
	\section{Bohr-Mottelson model with a minimal length }
	\label{sec:2}
	In the context of the collective geometrical  model of Bohr-Mottelson\cite{Bohr}, the classical expression for the rigid-body kinetic energy associated with the rotation and surface deformations of	a nucleus has the form \cite{Bohr,Sitenko}
	\begin{equation}
	\hat{T} = \frac{1}{2}\sum_{k=1}^{3} {\cal J}_k\, \omega^{\prime2}_k +
	\frac{B_m}{2}\,(\dot{\beta}^2+\beta^2 \dot{\gamma}^2),
	\label{E9}
	\end{equation}
	where $\beta$ and $\gamma$ are the usual collective variables, $B_m$ is the
	mass parameter. Also,
	\begin{equation}
	{\cal J}_k = 4B_m\beta^2 \sin^2\bigl(\gamma - {\textstyle\frac{2}{3}}\pi k\bigr)
	\label{E10}
	\end{equation}
	are the three principal irrotational moments of inertia,
	and $\omega^\prime_k$ ($k=1$, 2, 3) are the components of the angular velocity (angular frequencies)
	on the body-fixed $k$-axes,
	which can be expressed in terms of the time derivatives of the Euler angles
	$\dot{\phi}, \dot{\theta}, \dot{\psi}$ \cite{Sitenko,Zare}
	\begin{eqnarray}
	\omega^\prime_1 &=& -\sin\theta \cos\psi\,\dot{\phi} + \sin\psi\,\dot{\theta},
	\nonumber\\
	\omega^\prime_2 &=& \sin\theta \sin\psi\,\dot{\phi} + \cos\psi\,\dot{\theta},\\
	\omega^\prime_3 &=& \cos\theta\,\dot{\phi} + \dot{\psi}. \nonumber
	\label{E11}
	\end{eqnarray}
	Going further, by  assuming the nucleus to be $\gamma$-rigid (i.e. $\dot{\gamma}=0$), as a non-adiabatic approach proposed by Davydov and Chaban in \cite{Davydov1}, and considering in particular the axially symmetric prolate case of $\gamma=0$, we see that
	the third irrotational moment of inertia ${\cal J}_3$ vanishes, while
	the other two become equal ${\cal J}_1 = {\cal J}_2 = 3B_m\beta^2$, thus the kinetic
	energy of Eq. (\ref{E9}) is simply \cite{Sitenko,Davydov2}
	\begin{align}
	\hat{T} =& \frac{3}{2} B_m\beta^2 (\omega^{\prime2}_1 + \omega^{\prime2}_2)
	+ \frac{B_m}{2}\,\dot{\beta}^2\nonumber \\
	=& \frac{B_m}{2}\Bigl[3\beta^2(\sin^2\theta\,\dot{\phi}^2 + \dot{\theta}^2)
	+ \dot{\beta}^2 \Bigr].
	\label{E12}
	\end{align}
	Since in the case of axial symmetry the nucleus can rotate only about directions perpendicular to the symmetry axis,  the collective motions in the nucleus are characterized by only three degrees of freedom:  $q_1=\phi$, $q_2=\theta$, and $q_3=\beta$.	Having in mind the position space representation (\ref{E8}), the kinetic energy operator, in this case, can be expressed in terms of the Laplacian and bi-Laplacian operators as follows 	
	\begin{align}
	T=-\frac{\hbar^2}{2B_m}\left(1-2\alpha\hbar^2\frac{1}{\sqrt{g}}\sum_{ij}\frac{\partial}{\partial q_i}\sqrt{g}g_{ij}^{-1}\frac{\partial}{\partial q_j}\right)\nonumber\\ \times\left[\frac{1}{\sqrt{g}}\sum_{ij}\frac{\partial}{\partial q_i}\sqrt{g}g_{ij}^{-1}\frac{\partial}{\partial q_j}\right]
    \label{E13}
	\end{align}
		where the matrix $g_{ij}$ having a diagonal form
		\begin{equation}
			g_{ij} = \left(\begin{array}{ccc} 3\beta^2\sin^2\theta & 0 & 0 \\ 0 &
				3\beta^2 & 0 \\0 & 0 & 1 \end{array}\right)
				\label{E14}
		\end{equation}
		where $g$ is the determinant of the matrix $g_{ij}$ and $g_{ij}^{-1}$ is the inverse matrix of  $g_{ij}$.
	Using the general procedure of quantization (Pauli–Podolsky prescription) in curvilinear coordinates, we obtain, in compact form, the collective Hamiltonian operator, up to the first order of $\alpha$,
	\begin{align}
	\hat{H}=-\frac{\hbar^2}{2B_m}\Delta +\frac{\alpha\hbar^4}{B_m}\Delta^2+ V(\beta)
	\label{E15}
	\end{align}
	with
	\begin{align}
	\Delta = \Biggl[\frac{1}{\beta^2}
	\frac{\partial}{\partial\beta}\beta^2\frac{\partial}{\partial\beta} +
	\frac{1}{3\beta^2} \Delta _\Omega\Biggr]
	\label{E16}
	\end{align}
	where $\Delta_\Omega$ is the angular part of the Laplace operator
	\begin{equation}
	\Delta_\Omega = \frac{1}{\sin\theta}\frac{\partial}{\partial\theta}\sin\theta
	\frac{\partial}{\partial\theta} + \frac{1}{\sin^2\theta}
	\frac{\partial^2}{\partial\phi^2}.
   \label{E17}
	\end{equation}
	The corresponding deformed Schr\"odinger equation to the first order on $\alpha$ reads as
	\begin{equation}
	\Biggl[-\frac{\hbar^2}{2B_m}\Delta +\frac{\alpha\hbar^4}{B_m}\Delta^2+ V(\beta)-E \Biggr] \Psi(\beta,\theta,\phi)=0
   \label{E18}
	\end{equation}
	which is a second order differential equation.
	In addition, we can see, here, that is difficult to obtain analytic solution of this differential equation, because of the bi-Laplacian $\Delta^2\propto p^4$ . However, we can get rid of  the term $\Delta^2$  in equation (\ref{E18}) by introducing an auxiliary wave function $\Phi$ as in \cite{Problem9}, so that
	\begin{equation}
	\Psi(\beta,\theta,\phi)=\left[1-2\alpha(i\hbar)^2\Delta\right]\Phi(\beta,\theta,\phi)
	\label{E19}
	\end{equation}
	Thus, we obtain the following differential equation satisfied by $\Phi$,
	\begin{equation}
	\Biggl[\left(1+4B_m\alpha\left(E-V(\beta)\right)\right)\Delta+\frac{2B_m}{\hbar^2}\left(E-V(\beta)\right)\Biggr] \Phi(\beta,\theta,\phi)=0
	\label{E20}
	\end{equation}
	where $\Delta$ is defined by Eq. (\ref{E16}).
	 The  latter equation can be  solved by using the usual following factorization
	\begin{equation}
	\Phi(\beta,\theta,\phi) = F_{n_{\beta}}(\beta)\, Y_{LM}(\theta,\phi),
	\label{E21}
	\end{equation}
	where $Y_{LM}(\theta,\phi)$ are the spherical harmonics. Then
	the angular part leads to the equation
	\begin{equation}
	\Delta_\Omega Y_{LM}(\theta,\phi) =- L(L+1)Y_{LM}(\theta,\phi),
    \label{E22}
	\end{equation}
	where $L$ is the angular momentum quantum number,
	while  the radial part $F(\beta)$ obeys to:
	\begin{equation}
	\Biggl[\frac{1}{\beta^2}
	\frac{d}{d\beta}\beta^2\frac{d}{d\beta}
	-\frac{L(L+1)}{3\beta^2} + \frac{2B}{\hbar^2}\bar{K}(E,\beta)\Biggr]F_{n_{\beta}}(\beta) = 0.
	\label{E23}
	\end{equation}
	with
	\begin{equation}
	\bar{K}(E,\beta)=\left(\frac{E-V(\beta)}{\left(1+4B_m\alpha\left(E-V(\beta)\right)\right)}\right)
    \label{E24}
	\end{equation}
	and $n_{\beta}$ is the radial quantum number. Eq. (\ref{E23}) is an effective Schrödinger equation including the minimal length. It should be noticed that in the limit $\alpha \rightarrow 0$, Eq. (\ref{E23}) reduces to the ordinary collective Schrödinger equation \cite{Bonatsos1,Budaca1,Budaca2}.\\	
	In what concerns the $\beta$ degree of freedom, we will consider here an anharmonic
	behaviour reflected into an infinite square well shape of the potential as in the case of  X(3) symmetry \cite{Bonatsos1}:
	\begin{align}
	V(\beta) = \left\{ \begin{array}{lcl} 0, && \rm{if}\ \beta\leq\beta_{\omega} \\
	\infty, &&  \rm{if}\ \beta > \beta_{\omega} \end{array} \right. ,
	\label{E25}
	\end{align}
	where $\beta_{\omega}$ indicates the width of the well.
	In this case the wave function $F(\beta)$ is a solution of the equation
	\begin{equation}
	\Biggl[\frac{d^2}{d\beta^2} + \frac{2}{\beta}\frac{d}{d\beta}
	+ \Biggl(\bar{k}-\frac{L(L+1)}{3\beta^2}\Biggr)\Biggr]F_{n_{\beta}}(\beta) = 0
	\label{E26}
	\end{equation}
	in the interval $0\leq\beta\leq\beta_{\omega}$,
	where we introduced the reduced energies
	\begin{equation}
	\varepsilon=\bar{k}=\frac{2B_m}{\hbar^2}\cdot\frac{E}{(1+4B_m\alpha E)}
    \label{E27}
	\end{equation}
	 while it vanishes outside.
	Substituting $F_{n_{\beta}}(\beta)=\beta^{-1/2} f_{n_{\beta}}(\beta)$ in Eq.(\ref{E26}), one obtains the Bessel equation
	\begin{equation}
		\Biggl[\frac{d^2}{d\beta^2} +
		\frac{1}{\beta}\frac{d}{d\beta}
		+ \Biggl(\bar{k}^2-\frac{\eta^2}{\beta^2}\Biggr)\Biggr]
		 f_{n_{\beta}}(\beta) = 0,
	\label{E28}
	\end{equation}
	with
	\begin{equation}
		\eta=\left(\frac{L(L+1)}{3}+\frac{1}{4}\right)^{\frac{1}{2}},
	\label{E29}
	\end{equation}
	and the boundary condition being $ f_{n_{\beta}}(\beta_{\omega})=0$.
	The solution of Eq.(\ref{E26}), which is finite at $\beta=0$, is then given by
	\begin{equation}
	F_{n_{\beta}}(\beta)=F_{sL}(\beta) = N_{s,L}\,\beta^{-1/2}
		J_{\eta}(\bar{k}_{s,\eta}\beta),  \ s=n_{\beta}+1
	\label{E30}
	\end{equation}
	with $ \bar{k}_{s,\eta}=\chi_{s,\eta}/\beta_{\omega}$ and $\varepsilon_{s,\eta}=\bar{k}_{s,\eta}^2$,
	where $\chi_{s,\eta}$ is the $s$-th zero of the Bessel function of the first kind
	$J_{\eta}(\bar{k}_{s,\eta}\beta_{\omega})$. $N_{s,L}$ is a normalization constant  to be determined later.
	The corresponding spectrum is then
	\begin{equation}
		E_{s,L} = \frac{\hbar^2}{2B_m}\times\frac{\bar{k}_{s,\eta}^2}{1-2\hbar^2\alpha\bar{k}_{s,\eta}^2},\ \ \bar{k}_{s,\eta}= \frac{\chi_{s,\eta}}{\beta_{\omega}}
		\label{E31}
	\end{equation}
	In the above equation, the term $2\hbar^2\alpha\bar{k}_{s,\eta}^2$ is the correction due
	to the minimal length. Therefore, we conclude that the minimal
	length increases slightly the energy spectrum. In addition, the relative
	correction can be written as
	\begin{equation}
	\frac{\Delta 	E_{s,L}  }{	E_{s,L}^{0} }=\frac{2\hbar^2\alpha\bar{k}_{s,\eta}^2}{1-2\hbar^2\alpha\bar{k}_{s,\eta}^2}
	\label{E32}
	\end{equation}
where  $E_{s,L}^{0}=\lim_{\alpha \rightarrow 0}E_{s,L}$ .\\
	Essentially, the total wave function (\ref{E19}) can be written as
	\begin{align}
		\Psi(\beta,\theta,\phi)&=\left[1-2\alpha(i\hbar)^2\Delta\right]\Phi(\beta,\theta,\phi)\nonumber \\
	%	                      &=\left[1-2\alpha\hbar^2\left(\frac{1}{\beta^2}
	%	                      \frac{\partial}{\partial\beta}\beta^2\frac{\partial}{\partial\beta} +
	%	                      \frac{1}{3\beta^2} \Delta _\Omega\right)\right]F_{n_{\beta}}(\beta) Y_{LM}(\theta,\phi)\nonumber \\
		                      &=\left(1+2(i\hbar)^2\alpha\bar{k}_{s,\eta}^2\right)F_{n_{\beta}}(\beta) Y_{LM}(\theta,\phi)
    \label{E33}
	\end{align}
	Finally, we have	
\begin{align}
\Psi(\beta,\theta,\phi)
=N_{s,L}\left(1+2(i\hbar)^2\alpha\bar{k}_{s,\eta}^2\right)\beta^{-1/2}
J_{\eta}(\bar{k}_{s,\eta}\beta) Y_{LM}(\theta,\phi)
\label{E34}
\end{align}
Using the normalization condition of this function, we easy obtain the factor $N_{s,L}$ :
\begin{equation}
N_{s,L}=\frac{\sqrt{2}}{\beta_{\omega}\,J_{\eta+1}(\chi_{s,\eta})\left(1+2(i\hbar)^2\alpha\bar{k}_{s,\eta}^2\right)}
\label{E35}
\end{equation}
Having the analytical expression of the normalized wave function, one can readily compute the B(E2) transition probabilities. Nevertheless, it should be remarked that  the full normalized wave function does not change by introducing  the concept of minimal length. Therefore, the B(E2) transition probabilities, which are expressed as,
\begin{equation}
	B(E2; sL \to s'L') = \frac{1}{2L+1}
	\left|\langle s'L'||T^{(E2)}||sL\rangle\right|^2
	\label{E36}
\end{equation}
 also remain unchanged by this formalism  and are similar to those obtained in \cite{Bonatsos1} where  $T^{(E2)}_\mu = t\,\beta\,\sqrt{\frac{4\pi}{5}}\,Y_{2\mu}(\theta,\phi)$ is the quadrupole operator for $\gamma= 0$ and $t$ is a scaling factor.
Here, some remarks concerning  X(5)  with a minimal length concept are worth to be mentioned:
\begin{itemize}
\item (1) : notice that, in this case,
the same Eq. (\ref{E28}) occurs, but with $\eta=\left( {L(L+1)\over 3} +{9\over 4}\right)^{1/2} $.
\item (2) : As in the case of X(3) model , the concept of minimal length  has no effect on the B(E2) transition probabilities of X(5).
\end{itemize}
Besides, from the requirement that the wave function be symmetric with respect to the perpendicular plan to the symmetry axis of the nucleus and passing through its center, it follows that only even values of the angular momentum $L$ are allowed. Therefore no $\gamma$ bands appear in the present models as expected, because the $\gamma$ degree of freedom has been initially frozen to $\gamma=0$.

\setlength{\tabcolsep}{5.7pt}
\begin{table}[H]
	\caption{Typical energy levels ( ground state ) of the X(3)-ML and X(5)-ML models, normalized to the $2_{g}^+$ excited state energy  for different
		values of the parameter $\alpha$ with $\hbar=1$.}
	\label{table1}
	\begin{center}
		{\linespread{1.93}
			\footnotesize
			\begin{tabular}{|c|c|c|c|}
				\hline
				L&\multicolumn{3}{|c|}{X(3)-ML}\\
				\hline
				$0^{+}$&   0.000 & 0.000 & 0.000 \\
				$2^{+}$&   1.000 & 1.000 & 1.000 \\
				$4^{+}$&   2.445 & 2.455 & 2.465\\
				$6^{+}$&   4.234 & 4.274 & 4.315\\
				$8^{+}$&   6.348 & 6.448 & 6.551 \\
				$10^{+}$&  8.779 & 8.980 & 9.194\\
				$12^{+}$&  11.520 & 11.880 & 12.270\\
				\noalign{\smallskip}\hline\noalign{\smallskip}
				$\beta_{\omega}$& -- & 60.0& 60.0\\
				$\alpha$& 0 & 0.5& 1\\
				\hline
			
			\end{tabular}
			}
			{\linespread{1.93}
				\footnotesize
				\begin{tabular}{|c|c|c|c|}
					\hline
					L&\multicolumn{3}{|c|}{X(5)-ML}\\
					\hline
					$0^{+}$&   0.000 & 0.000 & 0.000 \\
					$2^{+}$&   1.000 & 1.000 & 1.000 \\
					$4^{+}$&   2.904 & 2.914 & 2.925\\
					$6^{+}$&   5.430 & 5.477 & 5.526\\
					$8^{+}$&   8.483 & 8.610 & 8.741 \\
					$10^{+}$&  12.03 & 12.290 & 12.570\\
					$12^{+}$&  16.04 & 16.530 & 17.050\\
					\noalign{\smallskip}\hline\noalign{\smallskip}
					$\beta_{\omega}$& -- & 60.0& 60.0\\
					$\alpha$& 0 & 0.5& 1\\
					\hline
					
				\end{tabular}
			}
			
				\end{center}
			\end{table}
			\setlength{\tabcolsep}{13.8pt}
			\begin{table}[H]
				\caption{The values of the free parameters used in the calculations.}
				\label{table2}
				\begin{center}
					{\linespread{1.6}
						\footnotesize
						\begin{tabular}{|c|c|c|c|c|c|}
							\hline
							Models&\multicolumn{2}{c|}{X(3)-ML}	&\multicolumn{2}{c|}{X(5)-ML}\\
							\hline
							$Nucleus$&$\alpha$&$\beta_{\omega}$&$\alpha$&$\beta_{\omega}$\\
							\hline
							$ {}^{150}$Nd & 0.961& 29.446& 0.184 & 67.308\\
							$ {}^{176}$Os & 0.421& 42.517& 0.000 & 64.670\\
							$ {}^{178}$Os & 0.444& 38.575& 0.649 & 75.614\\
							$ {}^{180}$Os & 0.999& 21.858& 0.000 & 56.102\\
							$ {}^{156}$Dy & 0.833& 50.763& 0.000 & 95.399\\
							$ {}^{154}$Gd & 0.654& 60.299& 0.233 & 65.648\\
							\hline
						\end{tabular}}
					\end{center}
				\end{table}
				\setlength{\tabcolsep}{4.3pt}
	\section{Model applicability and numerical results}
\begin{figure*}[tbph]
			\centering
			\rotatebox{0}{\includegraphics[height=50mm]{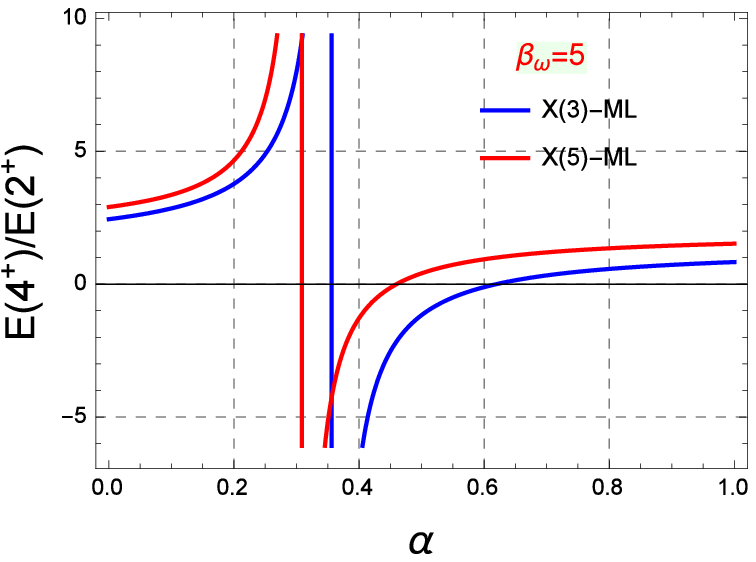}}
			\rotatebox{0}{\includegraphics[height=50mm]{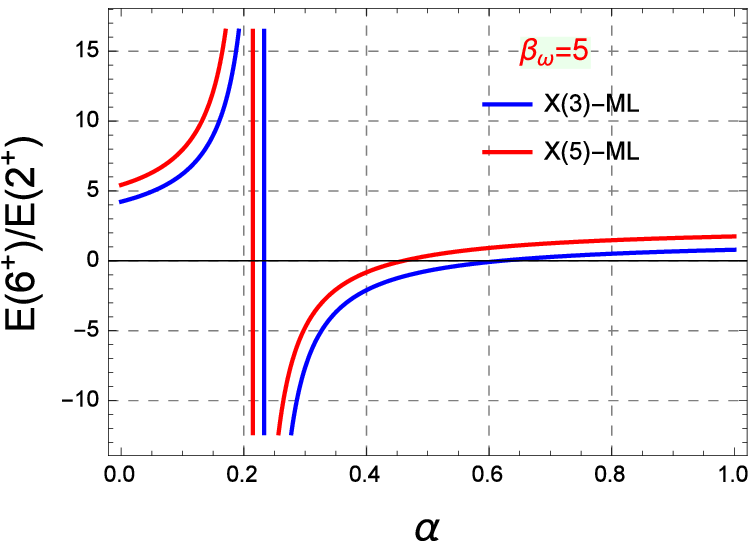}}
			\rotatebox{0}{\includegraphics[height=50mm]{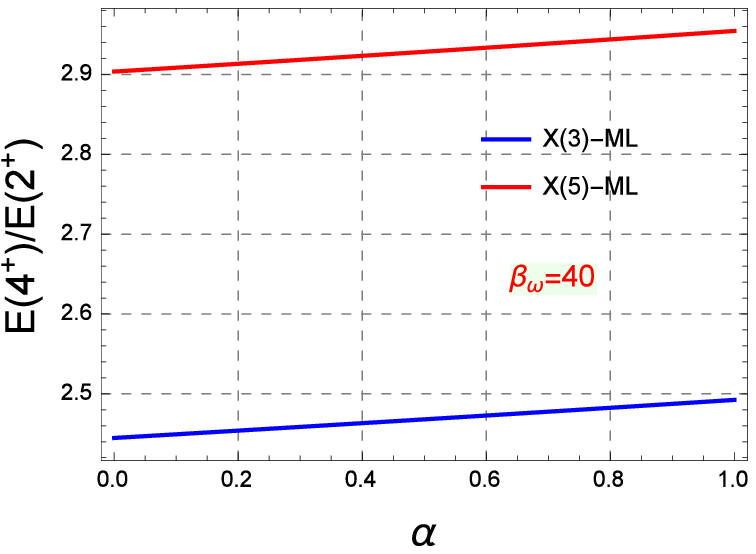}}
			\rotatebox{0}{\includegraphics[height=50mm]{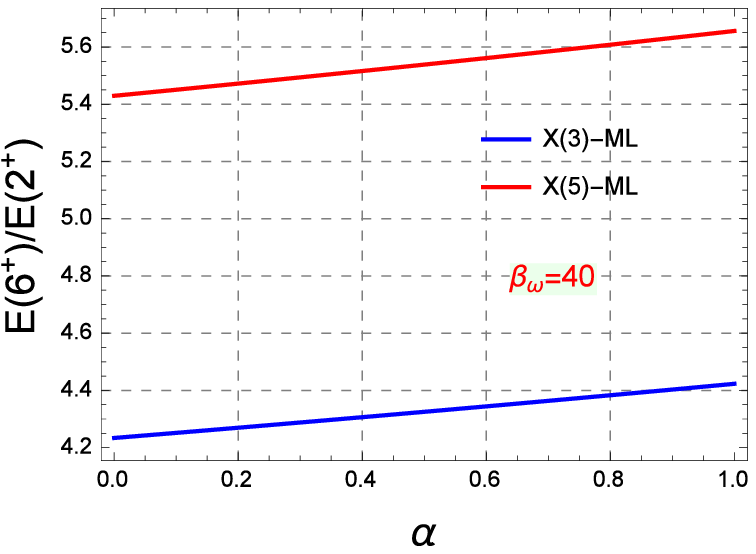}}
			\caption{The enegy of the first $4^+$ and $6^+$  levels are ploted as function of  the minimal length parameter $\alpha$. }
			\label{Fig1}
		\end{figure*}

\begin{figure*}[tbph]
		\centering
		\rotatebox{0}{\includegraphics[height=50mm]{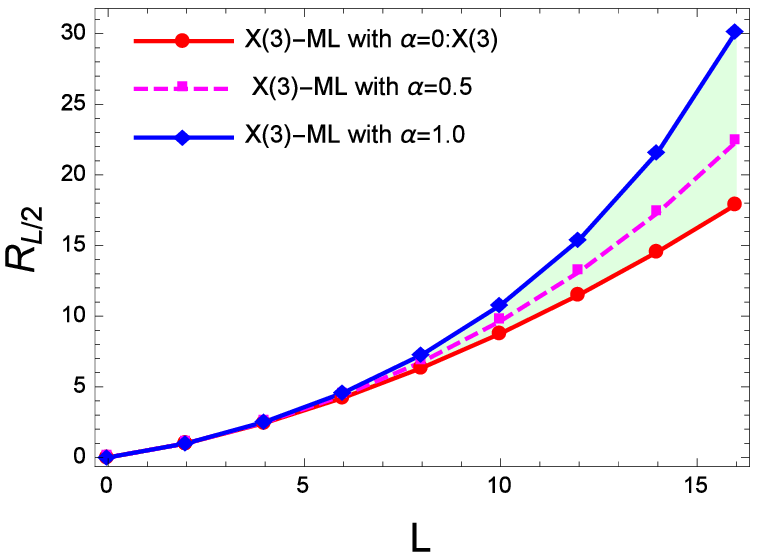}}
		\rotatebox{0}{\includegraphics[height=50mm]{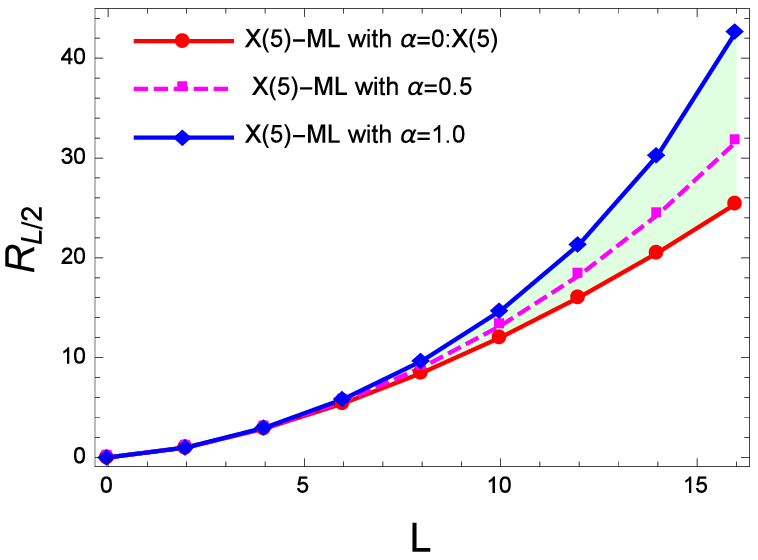}}
		\rotatebox{0}{\includegraphics[height=50mm]{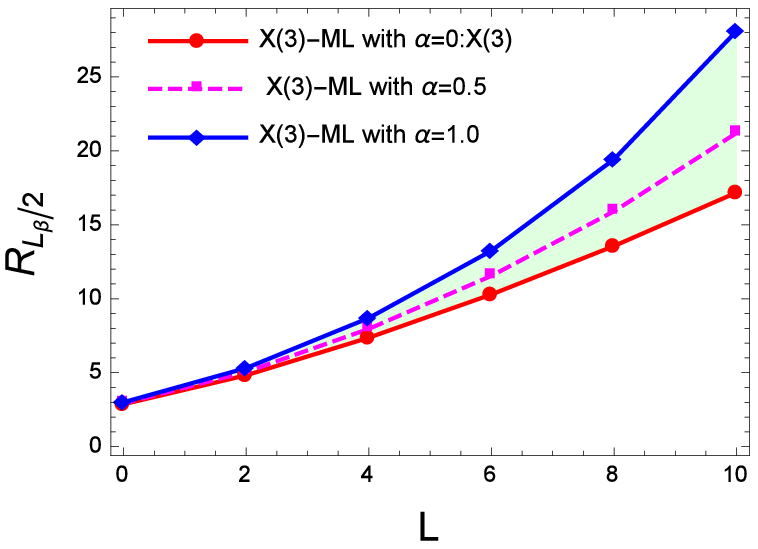}}
		\rotatebox{0}{\includegraphics[height=50mm]{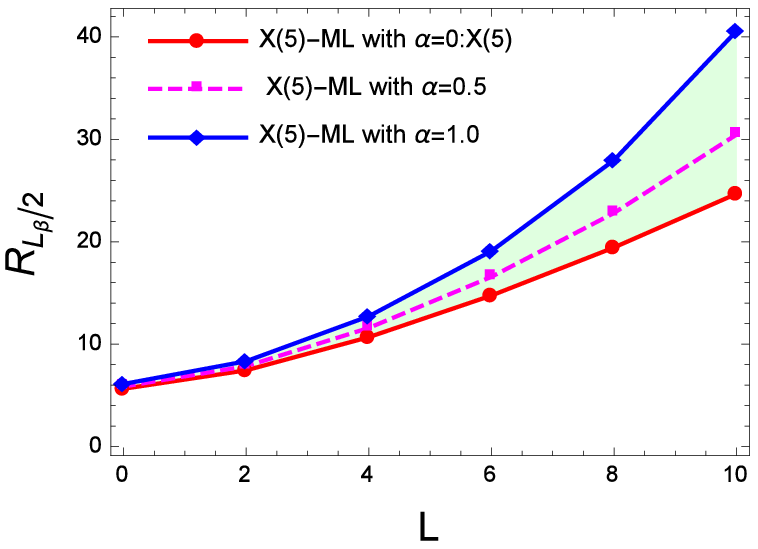}}
		\caption{The energy of the ground state  and the $\beta_1$ band, normalized to the energy of the first excited state in the X(3)-ML and X(5)-ML models are ploted as function of  angular momentum $L$ for different values of the minimal length parameter $\alpha$. The ground state is labeled by $R_{L/2}$, while $\beta_1$ band is labeled by $R_{L_{\beta}/2}$. The X(3)  and X(5) predictions are also shown for
			comparison. }
		\label{Fig2}
	\end{figure*}

\begin{figure*}[tbph]
	 	\centering
	 	\rotatebox{0}{\includegraphics[height=73mm]{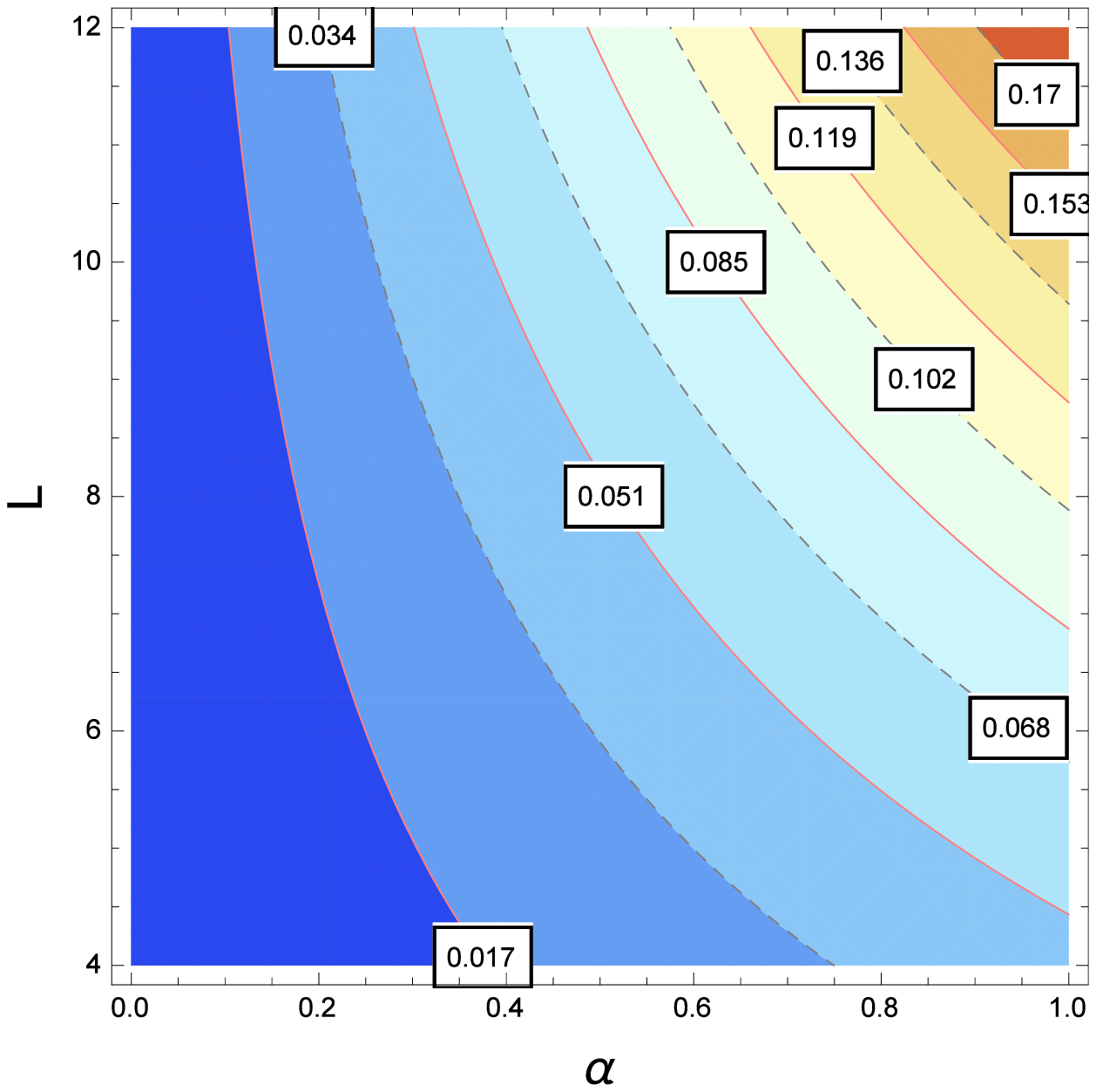}}
	 	\rotatebox{0}{\includegraphics[height=73mm]{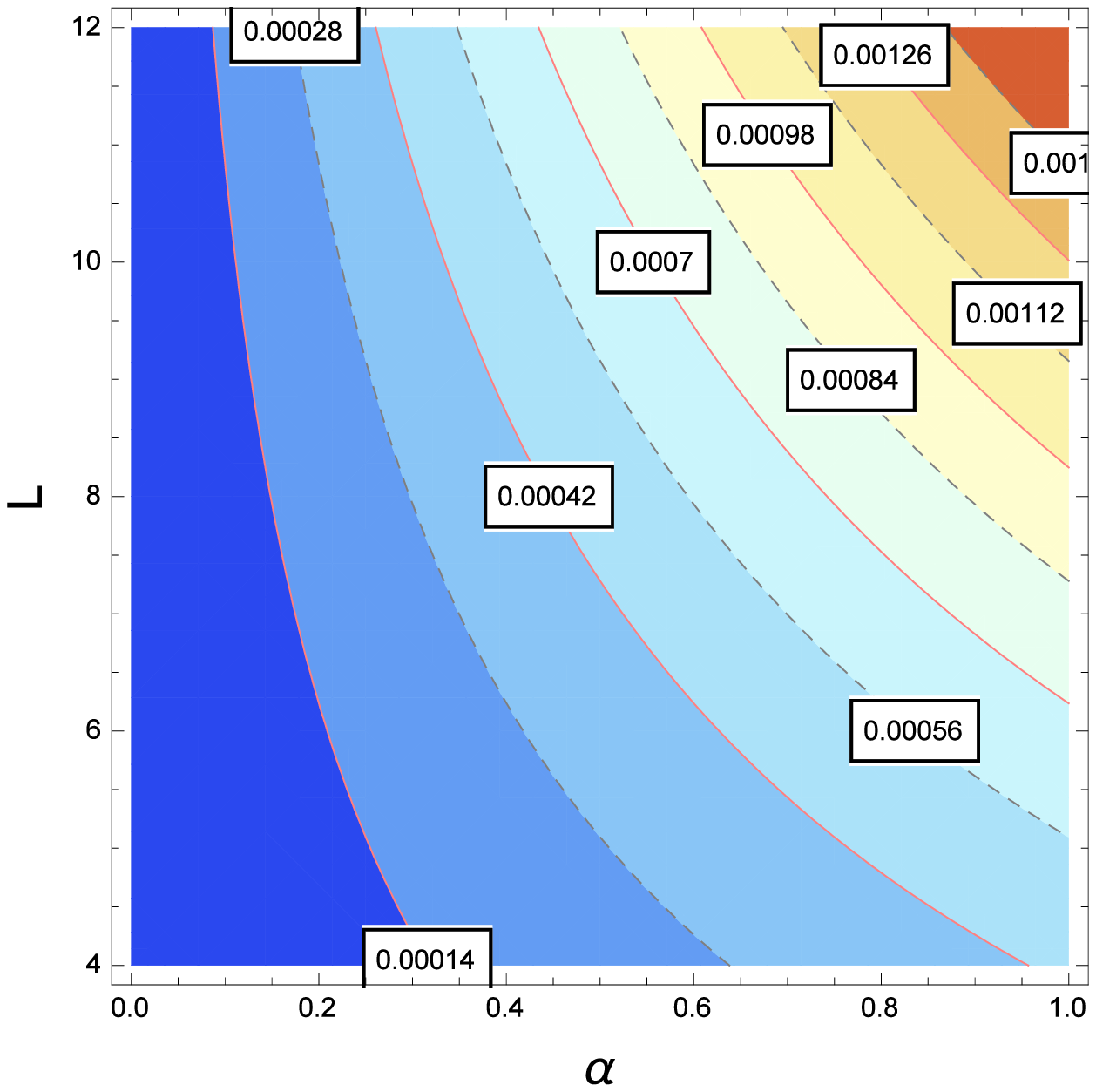}}
	 	\caption{Map contour lines of  the relative correction  (\ref{E32}) for the X(3)-ML model  drawn as a function  of  the angular momentum $L$ and the minimal length parameter $\alpha$ for $\beta_{\omega}=40$ (left) and $\beta_{\omega}=400$ (right).}
	 	\label{Fig3}
	 \end{figure*}

\begin{figure*}[tbph]
			\centering
			\rotatebox{0}{\includegraphics[height=30mm]{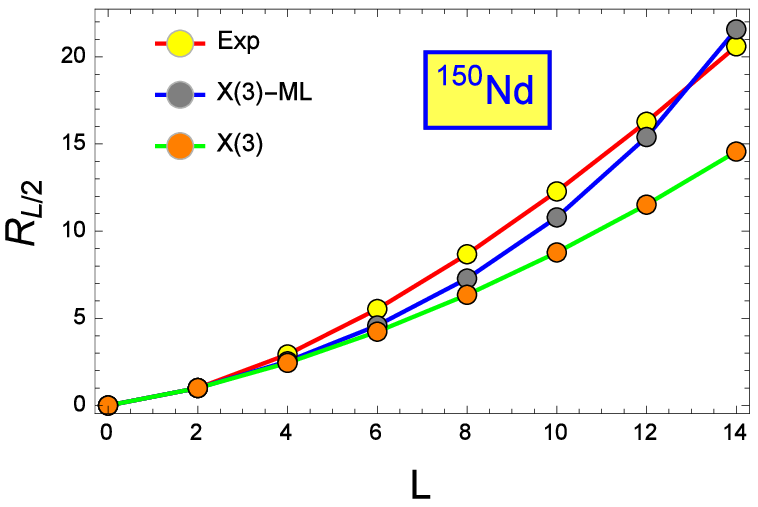}}
			\rotatebox{0}{\includegraphics[height=30mm]{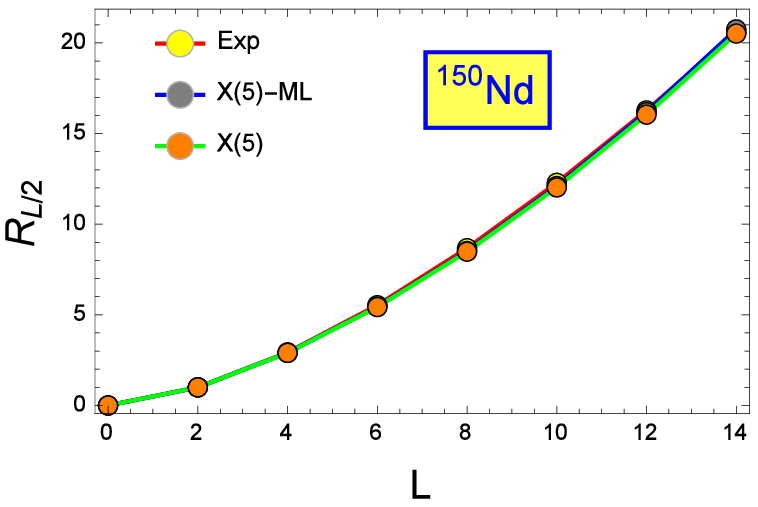}}
            \rotatebox{0}{\includegraphics[height=30mm]{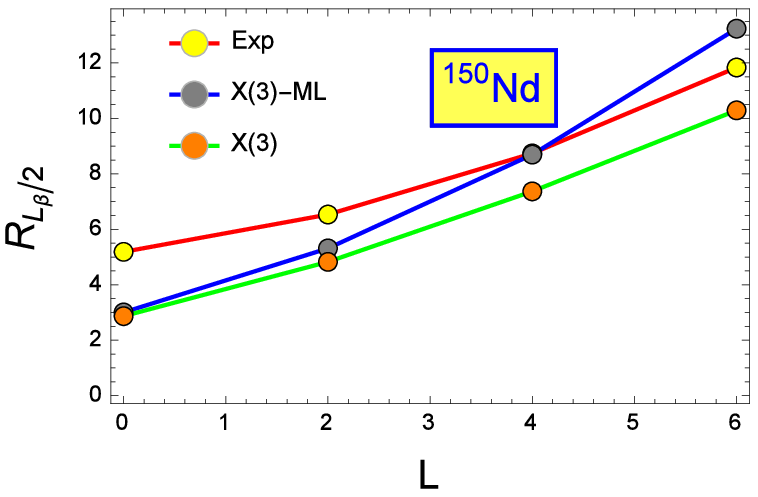}}
            \rotatebox{0}{\includegraphics[height=30mm]{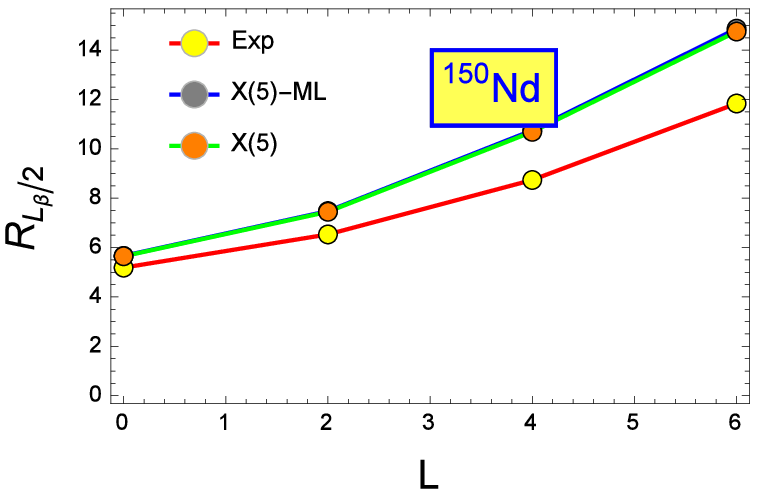}}
         	\rotatebox{0}{\includegraphics[height=30mm]{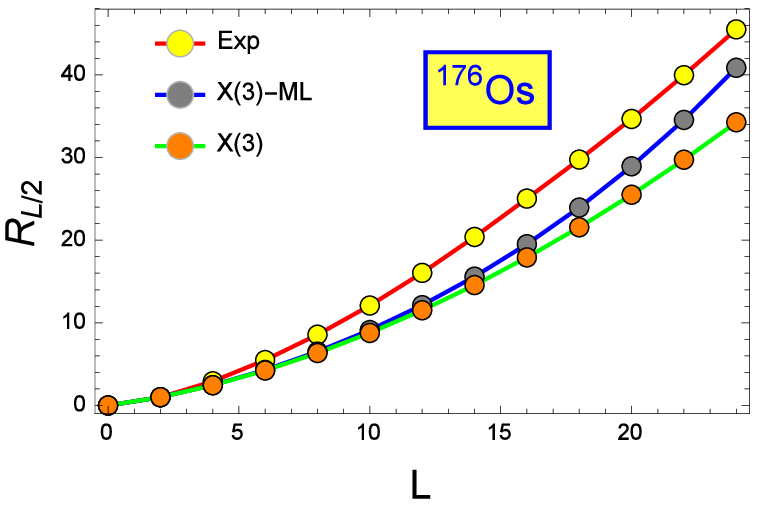}}
         	\rotatebox{0}{\includegraphics[height=30mm]{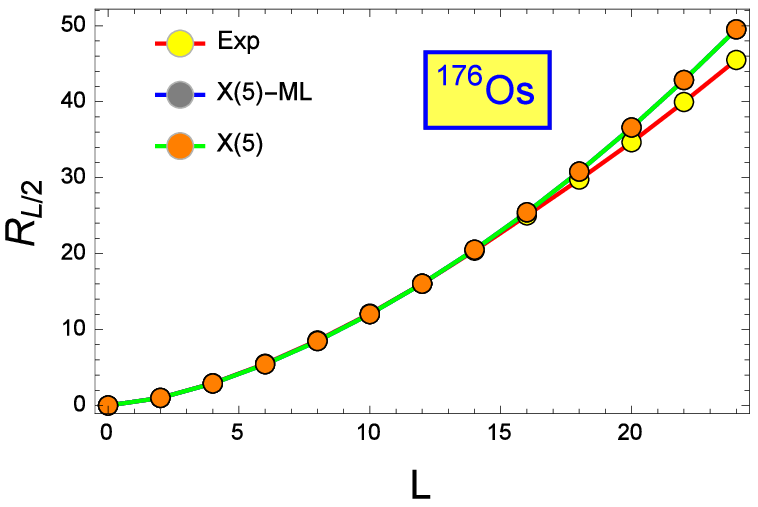}}
         	\rotatebox{0}{\includegraphics[height=30mm]{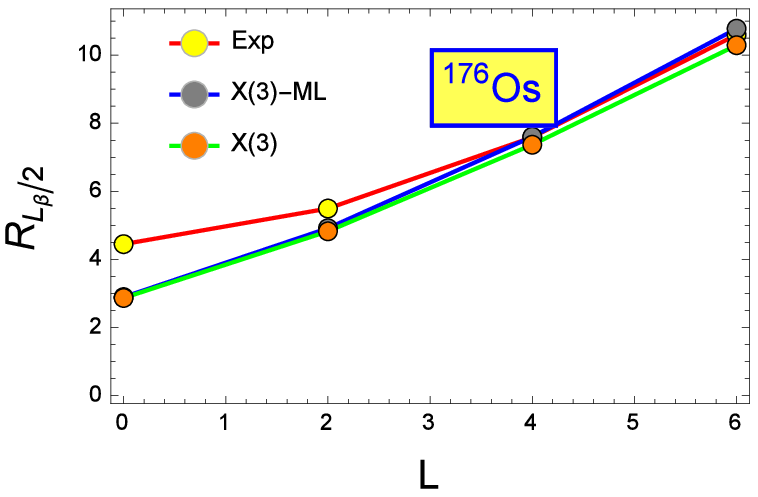}}
         	\rotatebox{0}{\includegraphics[height=30mm]{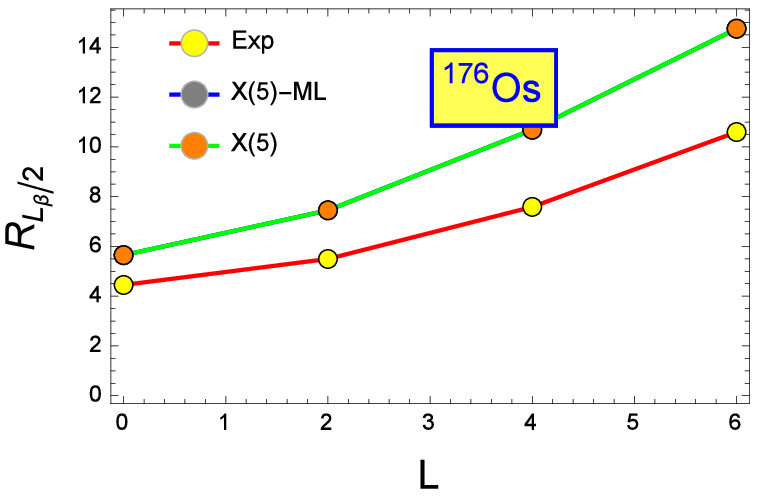}}
			\caption{Theoretical results for energy levels of the ground state and  the $\beta_1$-bands of the X(3)-ML and X(5)-ML  models, compared with the available experimental data \cite{dada} for ${}^{150}$Nd and ${}^{176}$Os. The levels of each band are normalized to the $2_1^+$ state. The ground state is labeled by $R_{L/2}$, while $\beta_1$ band is labeled by $R_{L_{\beta}/2}$.  }
			\label{Fig4}
		\end{figure*}

\begin{figure*}[tbph]
		 	\centering
		 	\rotatebox{0}{\includegraphics[height=30mm]{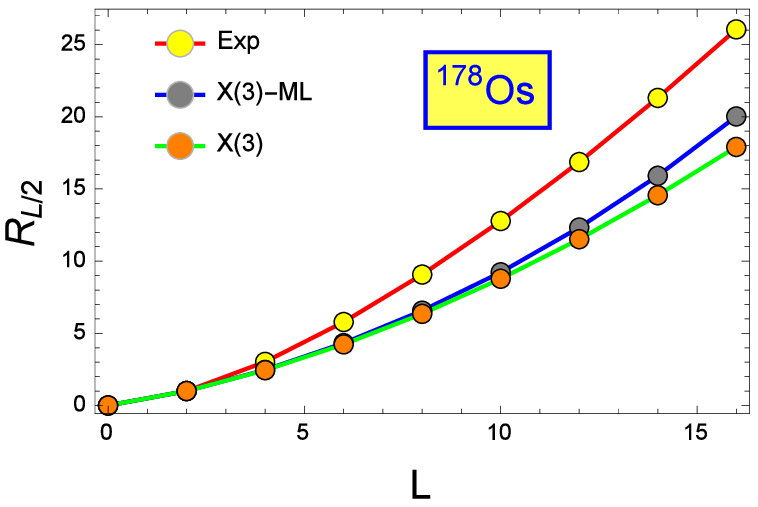}}
		 	\rotatebox{0}{\includegraphics[height=30mm]{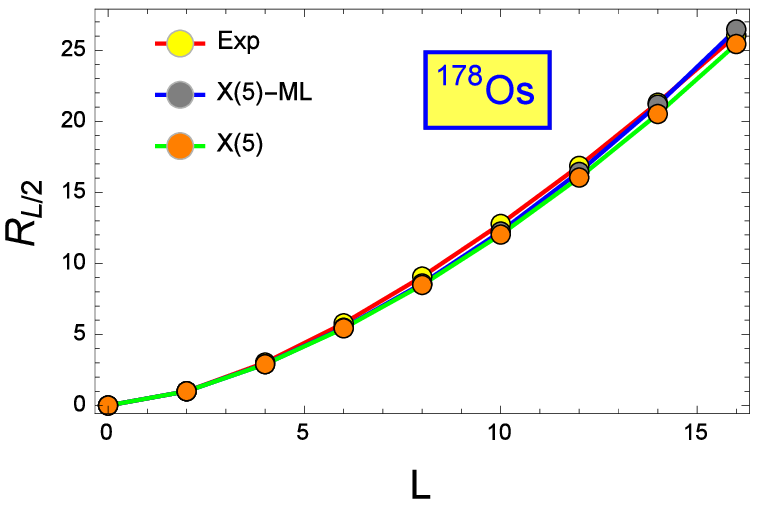}}
		 	\rotatebox{0}{\includegraphics[height=30mm]{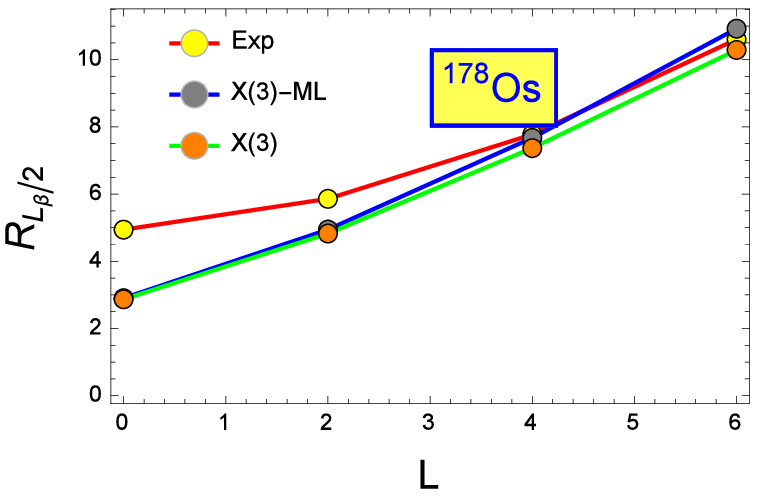}}
		 	\rotatebox{0}{\includegraphics[height=30mm]{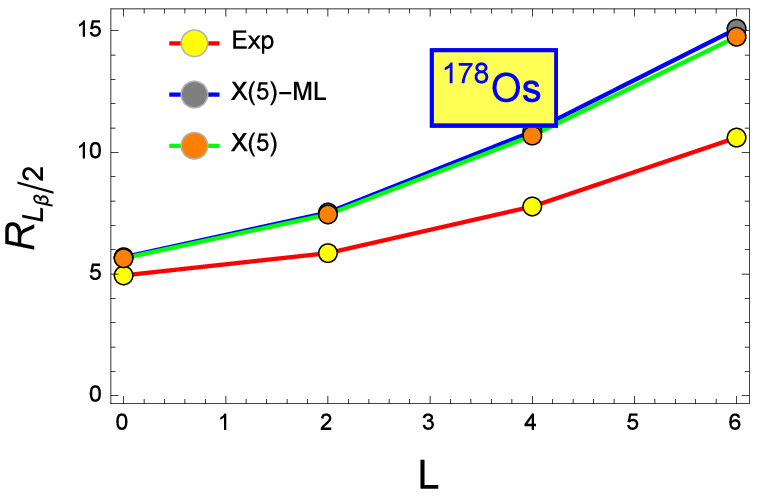}}
		 	\rotatebox{0}{\includegraphics[height=30mm]{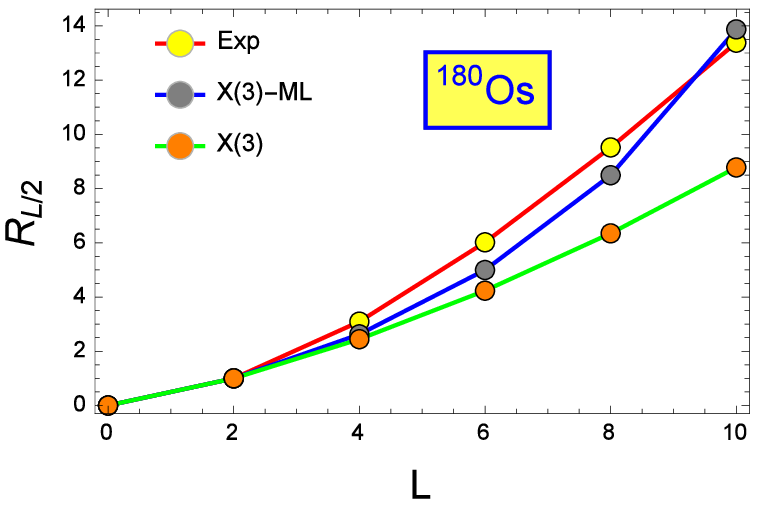}}
		 	\rotatebox{0}{\includegraphics[height=30mm]{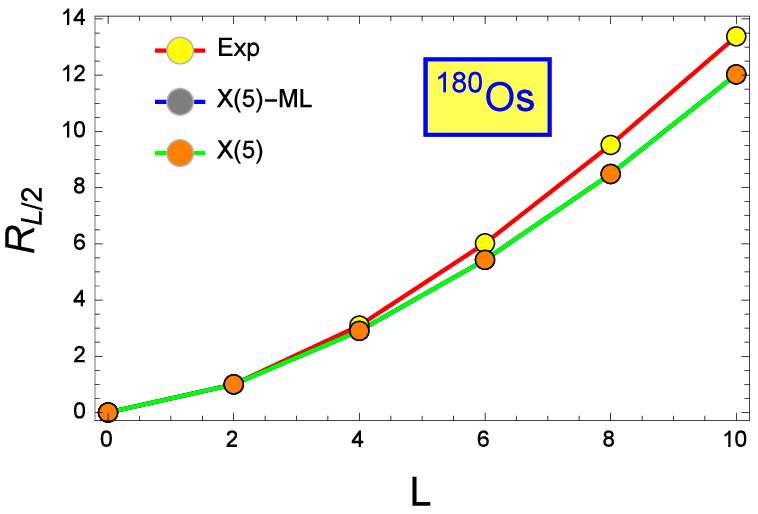}}
		 	\rotatebox{0}{\includegraphics[height=30mm]{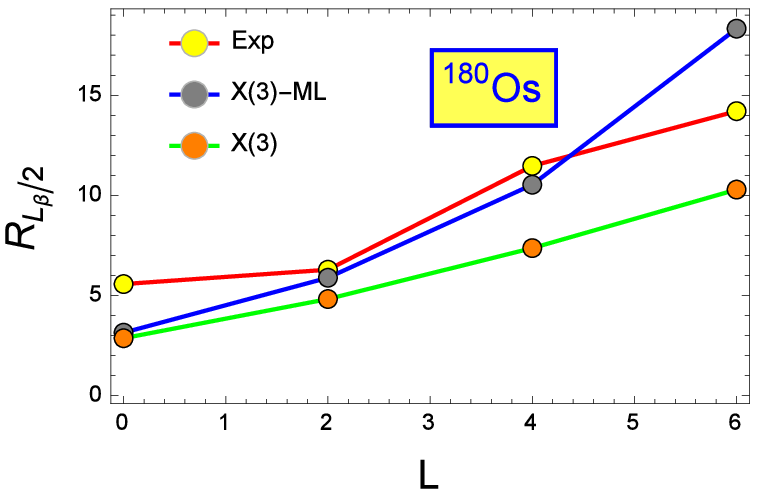}}
		 	\rotatebox{0}{\includegraphics[height=30mm]{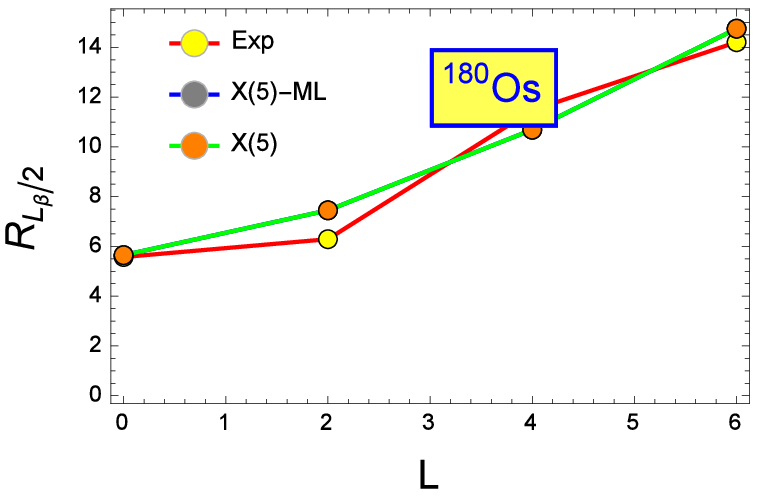}}
		 	\caption{Theoretical results for energy levels of the ground state and  the $\beta_1$-bands of the X(3)-ML and X(5)-ML  models, compared with the available experimental data \cite{dada} for ${}^{178}$Os and ${}^{180}$Os. The levels of each band are normalized to the $2_1^+$ state. The ground state is labeled by $R_{L/2}$, while $\beta_1$ band is labeled by $R_{L_{\beta}/2}$.  }
		 	\label{Fig5}
		 \end{figure*}

\begin{figure*}[tbph]
		  	\centering
		  	\rotatebox{0}{\includegraphics[height=30mm]{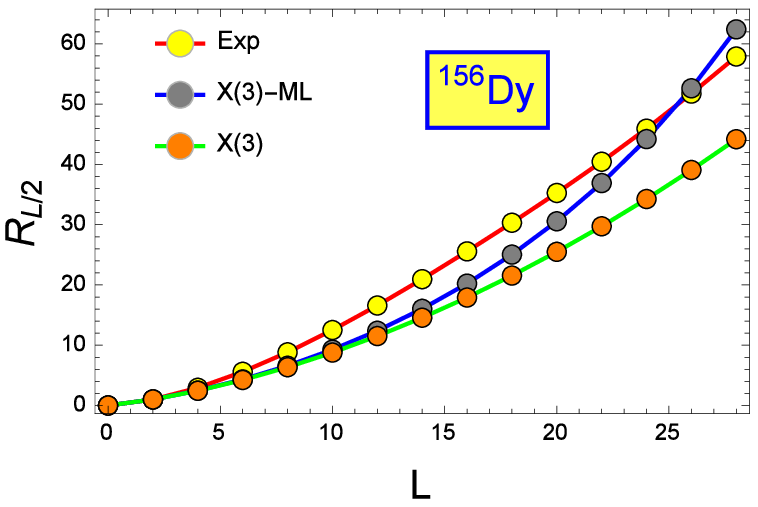}}
		  	\rotatebox{0}{\includegraphics[height=30mm]{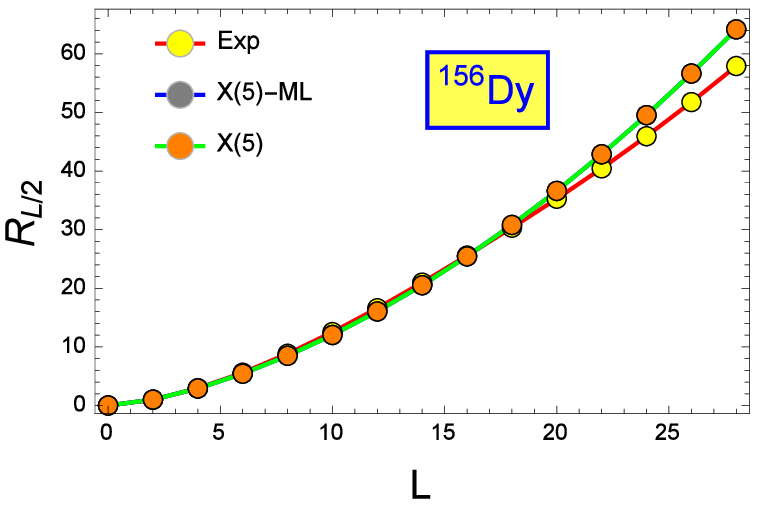}}
		  	\rotatebox{0}{\includegraphics[height=30mm]{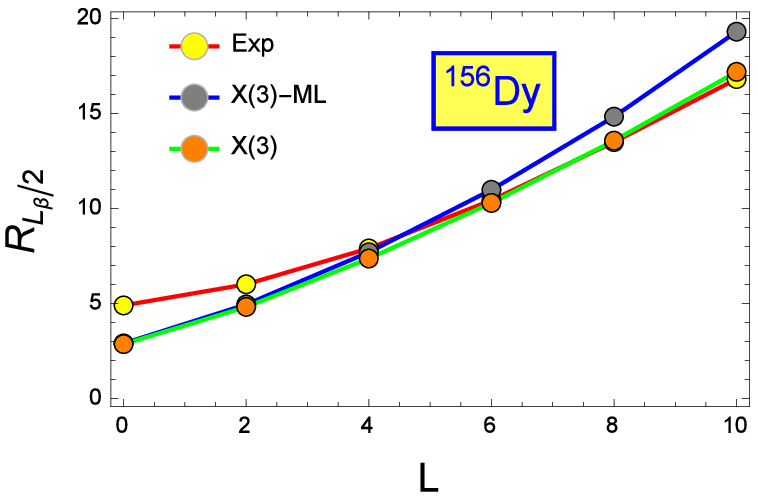}}
		  	\rotatebox{0}{\includegraphics[height=30mm]{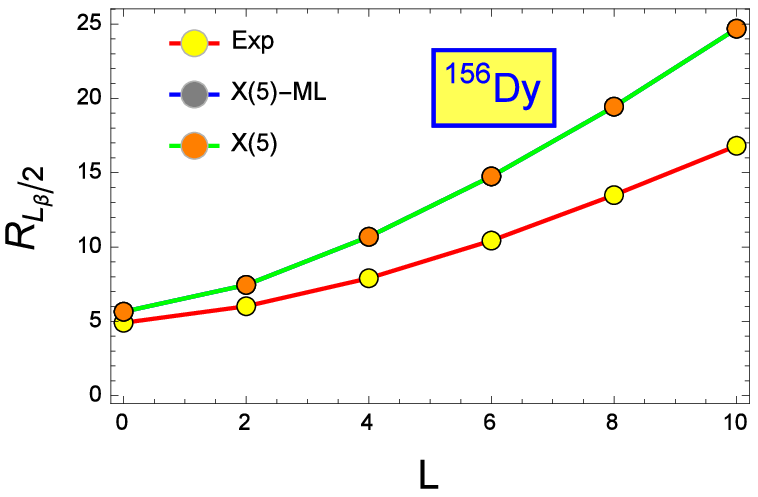}}
		  	\rotatebox{0}{\includegraphics[height=30mm]{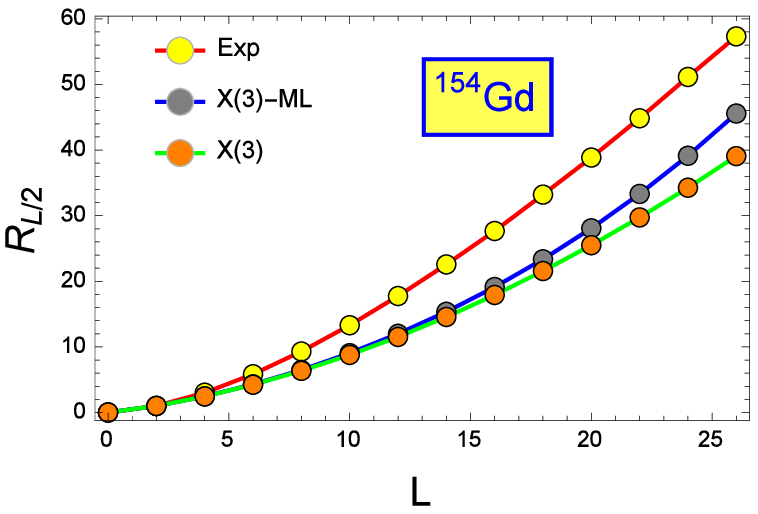}}
		  	\rotatebox{0}{\includegraphics[height=30mm]{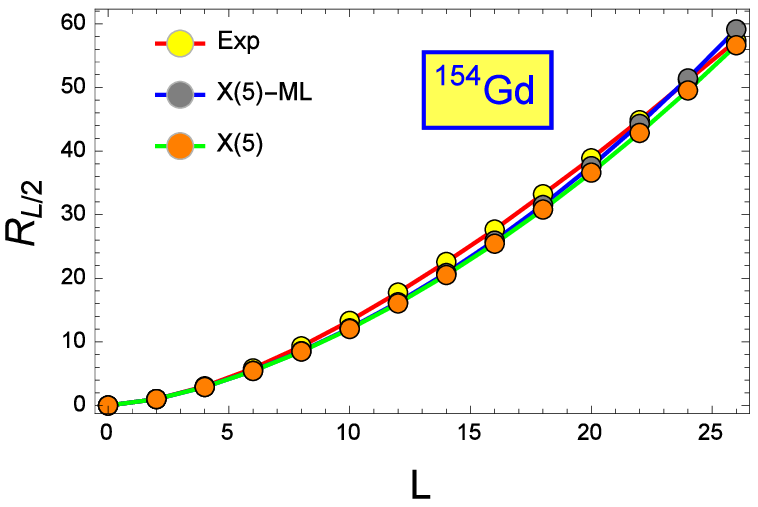}}
		  	\rotatebox{0}{\includegraphics[height=30mm]{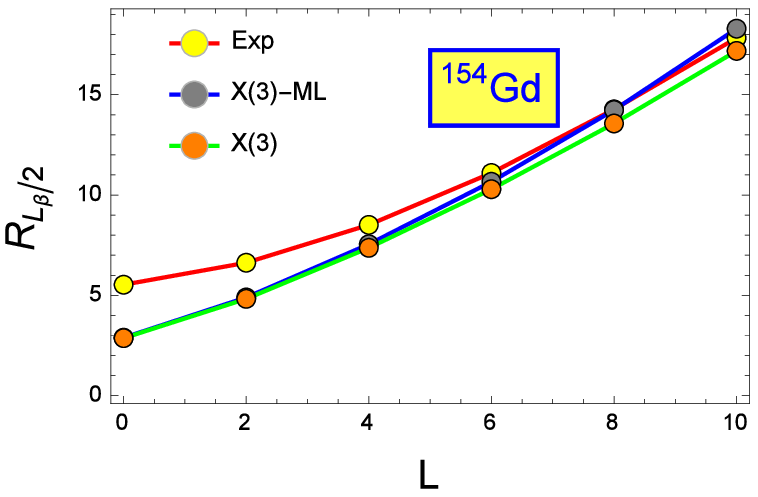}}
		  	\rotatebox{0}{\includegraphics[height=30mm]{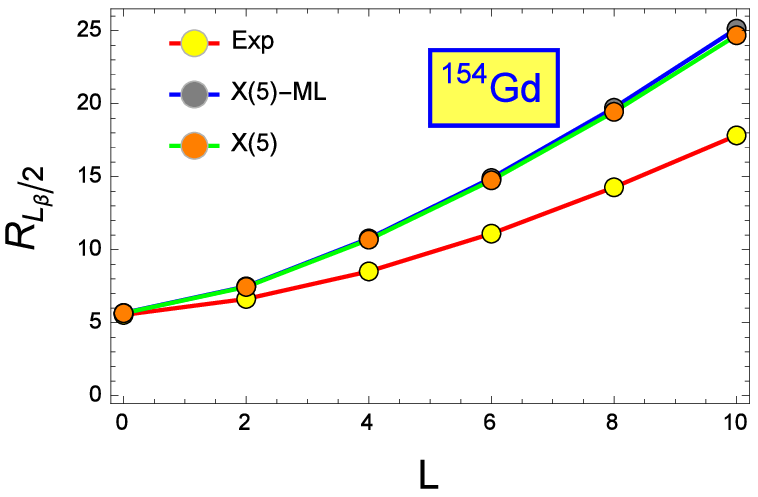}}
		  	\caption{Theoretical results for energy levels of the ground state  and  the $\beta_1$-bands of the X(3)-ML and X(5)-ML  models, compared with the available experimental data \cite{dada}  for ${}^{156}$Dy and ${}^{154}$Gd. The levels of each band are normalized to the $2_1^+$ state.  The ground state is labeled by $R_{L/2}$, while $\beta_1$ band is labeled by $R_{L_{\beta}/2}$.  }
		  	\label{Fig6}
		  \end{figure*}

Because the $\gamma$ degree of freedom has been frozen to $\gamma=0$, the bands in the present models, like in X(3) model, are only classified by the principal quantum number $n_{\beta}$ or $s=n_{\beta}+1$. A few interesting low-lying bands are given as
	\begin{itemize}
		\item i) The energy levels of the ground state band with $s = 1$,
		\item ii)  The $\beta$-vibrational bands with $s > 1$.
	\end{itemize}
In order to avoid any ambiguity of the nomenclature between our models and the existing phenomenological models, namely: X(3) and X(5) ,  we denote X(3)-ML and X(5)-ML  in connection with  X(3) and X(5) respectively.
 The proposed models have two free parameters, namely: the minimal length parameter $\alpha$ and the width of the infinite square well potential $\beta_{\omega}$. Obviously, we do not count the mass parameter $B_m$ since it disappears when calculating the energy ratios. However, according to the general form of the obtained energy spectrum, these parameters  could be dependent from each other and  check a constraint. Indeed, the energy  spectrum  corresponding to our models, where the effect of the minimal length  is considered, is always positive $E_{s,L}\ge 0$ (this is also valid in the ordinary case i.e: without a minimal length scenario). Due to this fact, we can write:
\begin{equation}
1-2\hbar^2\alpha\bar{k}_{s,\eta}^2>0, \ \bar{k}_{s,\eta}= \frac{\chi_{s,\eta}}{\beta_{\omega}}
\label{E37}
\end{equation}
which is a constraint  between $\alpha$ and $\beta_{\omega}$. From practical point of view, it is important to note that the value of $\alpha$ must  be very small compared to the width of the well $\beta_{\omega}$ in order to preserve the mentioned above constraint.
In Fig.~\ref{Fig1}, the energy  of the first $4^+$ and $6^+$  levels, of X(3)-ML and X(5)-ML models, for two values of the width $\beta_{\omega}=5$ and $\beta_{\omega}=40$, are displayed as function of the minimal length parameter $\alpha$ in the interval $[0, 1]$.
In the case of small value of $\beta_{\omega}=5$ , we see that, the energy  ratios of the first $4^+$ and $6^+$  levels, for X(3)-ML, present a singularity nearby  $\alpha=0.3561$ and $\alpha=0.2333$ respectively, because the condition (\ref{E37}), in this case, is not fulfilled. Likewise in the  case of X (5)-ML, but in this time, the singularity occurs around  the following values $\alpha=0.3087$ and $\alpha=0.2149$.
 While in the  case of a large value of $\beta_{\omega}=40$, where the above relationship is very well checked, the energy of the first $4^+$ and $6^+$  levels is very much influenced by $\alpha$ .
In addition, Table.~\ref{table1}  shows  a typical energy levels of ground state  of the X(3)-ML and X(5)-ML models, normalized to the $2_{g}^+$ excited state energy  for $\beta_{\omega}=60$.
%---------------------------------------------------------------------
From this table, one can see the effect of the minimal length becomes manifest for higher values of the angular momentum. Indeed, such a fact, which results from the uncertainty principle Eq. (\ref{E2}) as expected from string theory, is well illustrated schematically in Fig.~\ref{Fig2} where the evolution of the energy spectrum of the ground state and the $\beta_1$ bands, normalized to the first $2^+$ excited state, is presented. Furthermore, one can see that the effect of the minimal length is more important for the X (3) symmetry than for the X (5) one. Such an effect could be beneficial when trying to reproduce the experimental data for concrete nuclei in comparison, particularly, with the pure X (3) model as it can be seen subsequently. Moreover, from this figure one can see that the ground state band  as well as $\beta_1$ band are very much influenced by $\alpha$ for higher angular momentum. Besides, as is mentioned above, the minimal length effect increases slightly the energy spectrum.
%----------------------------------------------------------------------
  In Fig.~\ref{Fig3}, we present the variations of the relative correction of our model to the X(3) symmetry given by Eq. (\ref{E32}), as a function of the angular momentum L and  the minimal length  as well as the width $\beta_{\omega}$. The map contour lines are lines with a constant relative correction. The area delimited by two successive contour lines represents the recovery rate of the X(3) symmetry by our model. From Fig.~\ref{Fig3}, one can see that in the vicinity of  $\alpha\rightarrow 0$ and for lower values of the angular momentum L, the recovery area is large . So, in this region our model is identical to the X(3) one. But, as one goes in the same given region to higher values of L, such an area narrows. Also, as $\alpha$ increases, the recovery area starts to contract. So, the gap between our model and the X(3) one increases, as it was  mentioned above in the comment on Table~\ref{table1}. However, this gap between both models is worthwhile for ours insofar as it allows reproducing the experimental data, by our model, with a good precision in comparison with the pure X(3) model as it can be seen from Fig.~\ref{Fig4}, Fig.~\ref{Fig5} and Fig.~\ref{Fig6}.
  In the right panel of Fig.~\ref{Fig3}, given for $\beta_{\omega}=400$, we observe a similar behavior as in the left one for which $\beta_{\omega}=40$ but with a bit more contracted recovery areas and lower values of the relative correction corresponding to the contour lines. This is due to the fact that for a deeper square well, the minimal length becomes smaller in concordance with the constraint (\ref{E37}).
%----------------------------------------------------------------------
 As a result, the models, developed here, allow to describe properties of nuclei having the  signature $R_{4/2}=E(4_g^+)/E(2_g^+) \geq 2.44$, unlike the model developed in Ref \cite{Budaca1,Budaca2}  which studies a few  properties of nuclei having in this case the ratio $R_{4/2}< 2.44$.
The experimental realization of the models was found to occur in some nuclei $ {}^{150}$Nd,$ {}^{176-180}$Os,$ {}^{156}$Dy and $ {}^{154}$Gd, where the values of the used free parameters  in the calculations are listed in Table ~\ref{table2}.
%-------------------------------------------------------------------------
In Fig.~\ref{Fig4}, Fig.~\ref{Fig5} and Fig.~\ref{Fig6}, we present the numerical results for the energy of the ground state and the $\beta_1$ bands, normalized to the energy of the first excited level, obtained within X(3)-ML and X(5)-ML for these nuclei. From these figures, on can see the X(3)-ML model reproduces well the experimental data in comparison with the pure X(3) one in both bands. While, the X(5)-ML model is generally  identical to the X(5) one and slightly better in the case of ${}^{150}$Nd ,${}^{178}$Os and ${}^{154}$Gd nuclei. Such a difference, in the precision of predictions of both models: X(3)-ML and X(5)-ML, is due to the parameter $\eta$ which enters in the zeros of the Bessel function $\chi_{s,\eta}$ and so defining the energy through the quantity $\bar{k}_{s,\eta}$ Eq. (\ref{E31}).
Indeed, for a given L, the value of $\eta$  is lower in the X(3)-ML than in the X(5)-ML. Hence, the minimal length formalism seems to be more suitable for studying $\gamma$-rigid nuclei in the frame of the X(3) symmetry.
%-------------------------------------------------------------------------
			%--------------------------------------------------------------------------
	\section{Conclusion}
	In this work, we have derived new  solutions of the Bohr-Mottelson Hamiltonian in the $\gamma$-rigid regime  within the minimal length formalism which emerges in many higher dimension theories of quantum physics. The  recall potential of the collective $\beta$-vibrations is assumed to be equal to an infinite square well as in the standard X(3) and X(5) models. So, improved versions of the X(3) and X(5) symmetries being called X(3)-ML and X(5)-ML  are elaborated. Indeed, we have shown, through this work, that  the introduction of the minimal length formalism allows one to enhance the numerical calculation precision of physical observables, particularly the energy spectrum of nuclei in comparison with the X(3) and X(5) models. These later could be easily recovered by taking a null minimal length solutions of our models: X(3)-ML and X(5)-ML.
    \section*{References}
	\bibliographystyle{elsarticle-num}
	\bibliography{references}
\end{document}